To the National Research Council

*Committee on a Strategy Optimize the U.S. OIR System in the Era of the LSST*

- Principal Author (*Doe, John J.*): Liu, Charles T.
- Today's Date *(mm/dd/yy)*: 10/06/14
- Institution of Principal Author: College of Staten Island, City University of New York
- Co-Authors (*Doe, Jane J., Institution, etc.*): Willman, Beth, Haverford College, Pepper, Joshua, Lehigh University, Rutkowski, Michael, University of Minnesota, Norman, Dara, NOAO, Cruz, Kelle, Hunter College, Bochanski, John, Rider University, Lee, Hyun-chul, University of Texas – Pan American, Isler, Jedidah, Syracuse University, Gizis, John, University of Delaware, Smith, J. Allyn, Austin Peay State University, Moustakas, John, Siena College, Wehner, Elizabeth, University of St. Thomas, Alfred, Marcus, Howard University, McGruder, Charles, Western Kentucky University, Hoffman, Jennifer L., University of Denver, Kwitter, Karen, Williams College, Carini , Michael, Western Kentucky University, Bary, Jeff, Colgate University, Covey, Kevin R., Western Washington University, Finn, Rose, Siena College, Penprase, Bryan, Pomona College, Gelderman, Richard, Western Kentucky University, Schuler, Simon, University of Tampa
- E-mail Address: charles.liu@csi.cuny.edu

**Maximizing LSST's Scientific Return: Ensuring Participation from Smaller Institutions**

The remarkable scientific return and legacy of LSST, in the era that it will define, will not only be realized in the breakthrough science that will be achieved with catalog data. This Big Data survey will shape the way the entire astronomical community advances -- or fails to embrace -- new ways of approaching astronomical research and data. In this white paper, we address the template questions #4, 5, 6, 8 and 9, with a focus on the unique challenges for smaller, and often under-resourced, institutions, including institutions dedicated to underserved minority populations, in the efficient and effective use of LSST data products to maximize LSST's scientific return.

The LSST survey will allow faculty and students at **less financially endowed and non-R1 institutions** to engage in cutting-edge scientific investigations. Indeed, we believe that the full promise of LSST can be realized only if adequate attention is paid to meeting the research needs of scientists at smaller universities, colleges, and research centers. The number of departments offering a Bachelor of Science as their highest astronomy degree has nearly doubled over the last 9 years. Most of the increase is from smaller institutions that graduated about 20% of the students receiving an astronomy BS degree in 2012 (AIP Focus On Reports, 2013, http://www.aip.org/statistics/reports). That pattern of growth is likely to continue well into the LSST era, with additional growth coming from student populations historically underserved in scientific research fields. With adequate resources to support their unique needs, researchers at these colleges and universities are particularly suited to using LSST data both (1) to produce substantial scientific yield, and (2) to promote learning through scientific discovery, thanks to their approach to mentoring, teaching, and training a broad and diverse section of society for jobs both inside **and outside** of astronomical research.

What are these unique needs?

**Data Accessibility:** According to the 2013 AAS demographic survey, teaching (31%) and devising/conducting observations (25%) are the main professional activities of AAS members employed at 4-year colleges and universities. Only 6% cited data visualization/mining as their main activity, but it is these skills and alternative ways of approaching very large datasets that will be required with LSST. Growing the community of researchers -- especially in smaller and less financially endowed institutions -- with abilities in data mining, visualization and statistical analysis must be a core goal of the suite of resources available to support LSST science.

The Sloan Digital Sky Survey (SDSS) has generally been credited with revolutionizing the way astronomy research is conducted. A key reason has been the multiple ways in which SDSS data can be accessed, from point and click, to simple web forms, to complex web forms, to SQL queries that are well matched to super users. Over time, the SDSS portal has done an increasingly good job of communicating 'data nuances' to research users, e.g., how to select clean vs. complete data responsibly. This multi-faceted approach to accessibility will be important for LSST as well. However, the community cannot think of LSST science in the same way as SDSS science -- e.g., "write an SQL query; download data; analyze the data." This approach will not be effective for many LSST queries, and the analysis code will likely be too hardware intensive to be supported at smaller institutions. It may therefore be the case that such queries and analysis tools will need to be run at supercomputing centers that are managed by the major national partners off the small-institution investigator's site.

**Training/User Support:** While providing 'computing centers' will be crucial for researchers at smaller institutions, equally important will be resources to support training in the use of these query, analysis and visualization tools. At smaller institutions, the number of faculty in astronomy is typically small and curricular options are necessarily limited. The advent of major surveys like LSST requiring expertise in astroinformatics and astrostatistics means that small institutions are squeezed even more than larger ones in their ability to train students properly in the necessary skills with existing teaching resources. Thus, user support and training modules from the LSST project will be critically important.

A variety of options should be publicly available for researchers and students to become knowledgeable about tools and resource availability and use. These 'interpretation resources' should include online tutorials, sample queries, workshops, traditional helpdesk support as well as moderated online forums where users can share information. Such resources should be coordinated and maintained by LSST-expert support staff. Training must include details not only about how the data were obtained and processed but also information and informed recommendations about the proper ways to interpret large datasets and data-subsets using statistical tools, tasks and methods that allow researchers to derive meaningful science. These 'training resources' should start to be developed and made available pre-LSST, supporting data from current large surveys, in order to maximize their efficient use post-LSST.

**Access to Funding:** Opportunities for 'course buy-out' (teaching release) time, in addition to buying hardware and software infrastructure are particularly important for scientists at less financially endowed, higher teaching load institutions. Such grants would enable PIs to (for example) work at partner institutions to be trained on new techniques and tools for maximizing return on LSST data. Additional, but separate, grants to support LSST follow-up observing runs on ground-based facilities would further increase the participation of less endowed scientists in our O/IR system. A perpetual ground-based observing problem for smaller school researchers is the separation of money from ground-based observing which may leave researchers with great ideas unable to pursue them because of travel costs.

**Access to Follow-up Facilities:** Although significant science will be done with only the LSST catalog data, it is inevitable that many LSST discoveries will require follow-up resources, including access to telescopes, a variety of instrumentation and follow-on surveys at optical and other wavelengths. LSST follow-up and supporting observations on targeted facilities will be necessary for optimum scientific productivity in the LSST era. The southern hemisphere is a great match to LSST follow-up, but travel there is too expensive, both in money and time, for most researchers at small institutions with heavier teaching schedules. Options for access to telescopes, via remote or queue observing, should be actively explored and supported. The community should seize cost-effective opportunities to retain some **public** access to 4m+ facility(/ies) in the North. For example, NSF could prioritize shared nights on targeted facilities in subsequent MSIP calls, along the lines of TSIP in the past. As smaller telescopes (<4m) are privatized, they become less accessible to researchers at smaller institutions. The possibility of public time buy-in with public-private partnerships on smaller telescopes should remain an option as much as possible.

**SUMMARY:** LSST will lead a growing cultural shift in the approach to astronomical research. A greatly important part of this shift is a significant broadening of the types of institutions -- especially smaller and less financially endowed institutions -- where the scientific return of LSST will be realized. The recognition that those institutions have distinct needs, and the effort to develop the infrastructure and resources to help fulfill those needs, will thus be required in order to enable LSST to fulfill its key goals of "maximizing the usability of the data" and "worldwide participation in all data products" (LSST Science Book, version 2, 2009, p 9).